# Wetting behavior of supercooled water droplets impinging on nanostructured graphite surface: A molecular dynamics study

Amir Afshar and Dong Meng

Mississippi State University, Starkville, MS, 39762, USA

**In this work, we studied the wetting behavior of impinged and stagnant supercooled water nanodroplet at the atomistic scale using molecular dynamics simulations. We show that water droplet represents a retraction behavior from surface groove nanostructure after indicating maximum penetration. There is a threshold value for the value of impinging velocity, v*, below which penetration is not observed which depends on size of the surface roughness. This indicates grooves smaller than a certain size can stop water from penetration, considering v~0.5M in practical situations, which is such a useful information for design of nanostructured ice-phobic substrates. The resistance presumably comes from interfacial tension between solid substrate and liquid water droplet. The extent of water penetration significantly depends on droplet impinging velocity and the size of surface roughness, for impinging velocities higher than V*. Our result indicates the dominant role of droplet temperature on time required droplet recede from surface groove structure rather than droplet impinging velocity and the surface groove nanostructure.**

## 1. Introduction

Water droplets spreading and impacting on a solid surface represent substantial dynamic wetting phenomena in which there is an intersection between physics, chemistry and mechanics. They are pertinent for engineering applications such as surface coating and inkjet printing[1]. While static wetting is fairly well explained by Young's contact angle equation, the dynamics of water droplet wetting is more intricated, since the involved physical processes in the local region near the moving contact line become significant[2,3]. The wetting behavior of water droplet on solid surfaces is thoroughly understood on the macroscopic scale[4,5]; however there are many other new wetting phenomena happened at the nanoscale which is entirely different from the behaviors at the micro- and mesoscales. For numerous nanoscale applications the wetting of fluids on different surfaces, such as graphite[6] and metals[7] has been investigated.

Theoretically, for an inclusive understanding of the impacting and spreading process of water droplet on solid surface, some significant physical property name as droplet viscosity, inertial effect and surface tension between water droplet and solid surface needs to be considered. In terms of numerical modeling in approaching dynamic wetting some different methods such as volume of fluid (VOF)[8] and finite elements modeling (FEM)[9] have been used. These frequently used numerical methods provide information of the wetting process at the mesoscale level. In this work, we employed molecular dynamics (MD) simulation method which delivers information of dynamic wetting at the atomistic level. To the best of our knowledge, there have so far been no MD simulation incorporated to investigate the dynamic wetting of impinged water droplet on solid surfaces which has a numerous application in aeronautics, wind turbines and solar panels industries to name a few. At the nanoscale, MD is the best computational method to examine and predict the dynamic wetting properties of nanodroplets on solid surfaces[10].

On the nanoscale, the wetting property of a solid surface is significantly investigated by the atomistic interactions between the liquid and the surface. Specifically, the predication of wettability for a system consisting liquid droplet and solid surface, necessarily requires some detailed system information but not limited to surface structure, system temperature and chemical composition, which is implicitly contained in the atomistic simulation of materials. Thus, MD is an appropriate simulation tool to explore the nature of the surface wetting resulting from the interactions between liquid droplet and solid surface at the atomistic level. There are some works used MD simulations to study the wetting behavior of nanodroplet on various surfaces. Sedighi et. al.[11] and Park et. el.[12] studied the effect of strength of interfacial energy with respect to the droplet-surface and found increasing the interaction energy resulted in significant enhancement on the wettability.

The main objective of the present work is to investigate the dynamic wetting behaviors of impinged supercooled water nanodroplet on textured graphite surfaces using MD simulations. For this, the effect of system physical properties namely as surface roughness, droplet temperature and impinging velocity on dynamic wetting are also examined. The results are discussed in terms of evolution of depth of water penetration, droplet retracting time from surface and the maximum depth of penetration in which droplet experienced during dynamic wetting.

## 2. Model and simulation method

In this study, we modeled water molecules using TIP4P/Ice atomistic water model[13]. In comparison with other transferrable atomistic water models like as TIP4P/2005[14], TIP4P[15], TIP3P[16] and SPC[17], this water model appropriately reproduces structure and phase transition temperature of ice. It accurately reproduces transition temperature of 269.8$\pm$1K[18] at 1 bar and density of 0.906 g/cm$^3$ for hexagonal ice which is favorably comparable to the corresponding experimental values. Thus, this model is suitable for investigation of wetting behavior of supercooled liquid water on nanotextured substrate. Interaction parameters of TIP4P/ice force field are listed in Table 1 where $d_{OM}$ is the distance between the oxygen atoms and the massless charged point in the water model, and $q_H$ is the electron charge carried by each hydrogen atom in the water molecules.

Table 1. Interaction parameters of the water model[13]

| **Water model** | $\boldsymbol{\varepsilon_{O-O}}$**[kcal mol$^{-1}$]** | $\boldsymbol{\sigma_{O-O}}$**[Å]** | **$q_H$ (e)** | **$d_{OM}$ [Å]** |
|---|---|---|---|---|
| **TIP4P/Ice** | 0.21084 | 3.1668 | 0.5897 | 0.1577 |

Table 2. Lennard-Jones parameters for aromatic carbon atoms from OPLS force field[19]

| **Graphene carbon** | $\boldsymbol{\sigma_{C-C}}$**[Å]** | $\boldsymbol{\varepsilon_{C-C}}$ **[kcal mol$^{-1}$]** |
|---|---|---|
| | 3.55000 | 0.07 |

In our study, we introduced graphite blocks as the substrate. It consists of several graphene pillars with interlayer distance of 0.335nm. Each graphene sheet is made by several carbon atoms in a honey comb molecular structure which are bonded to their nearest neighbors at the equilibrium bond length and angle of 0.1418nm and 120$^\circ$ respectively[20]. All graphene sheets are electronically neutral during our simulation and are extended along 'z' direction of simulation box. They are modeled using OPLS (Optimized Potential for Liquid Simulation) force field with non-bonded interaction parameters given in Table 2.

Interaction between water molecules and the nanotextured graphite substrate at atomistic level are considered through the Lennard-Jones (LJ) non-bonded interaction potential existing between oxygen and carbon atoms in our system. Equation (1) shows the LJ potential governs the interactions existing in the system. In our study, we adopted energy well depth (ε) and the inter-particle distance (σ) between oxygen and carbon atoms by the Lorentz-Berthelot mixing rule[21] given by Equation (2) and (3), which we employed in our fracture work[22] and also the corresponding values reported by Werder et. al (2003)[6].

Table 3 provides the values of non-bonded interaction potentials between oxygen and carbon atoms in our system which are interacted by LJ potential.

$$U_{CO} = 4\varepsilon_{c-o}\left[\left(\frac{\sigma_{c-o}}{r}\right)^{12} - \left(\frac{\sigma_{c-o}}{r}\right)^{6}\right] \quad (1)$$

$$\sigma_{c-o} = \frac{1}{2}\left(\sigma_{c-c} + \sigma_{o-o}\right) \quad (2)$$

$$\varepsilon_{c-o} = \sqrt{\varepsilon_{c-c} + \varepsilon_{o-o}} \quad (3)$$

All simulations performed using the open-source molecular dynamics (MD) parallel simulator code, LAMMPS[23]. Long-range electrostatic interactions between water/ice molecules are calculated using particle-particle particle-mesh (PPPM) method[24]. Water molecules are treated as the rigid body by keeping the bond length and bond angle fixed using the SHAKE algorithm[25]. In our study, graphene pillars are kept fixed in their lattice positions throughout the simulations. It has been verified that keeping graphene sheets fixed during the simulation does not influence the wetting behavior of supercooled liquid water on graphite substrate; though it remarkably reduces the computational cost[6,26].

In this study, equilibrated structure of water droplet is obtained by supercooling a slab of pre-equilibrated ice Ih[27] at different supercooling temperatures of 250K and 265K. The modeled hexagonal Ih ice consists of 8640 number of water molecules has a prismatic face and surface area of $6.6 \times 9.1$ nm$^2$. This face was brought in contact with graphite substrates during the wetting simulation. The thickness of the ice cube is 4.9 nm being perpendicular to the ice-substrate contact area.

To study the wetting behavior of supercooled liquid water droplet on nano-textured graphite substrate, we initially tuned the hydrophobicity of graphite substrate by measuring the equilibrium contact angle of liquid water droplet on graphene sheets. To investigate the equilibrium contact angle of liquid water droplet on graphene sheets, we placed the pre-equilibrated *I*h ice in 2.0nm perpendicular to five horizontal graphene sheets. Following that, ice cube was melted under NVT constraint at 300K for 2.0ns. Figure 1 shows a snapshot from a molecular structure of the constructed system having equilibrated Ih ice and water droplet on top of graphene sheets.

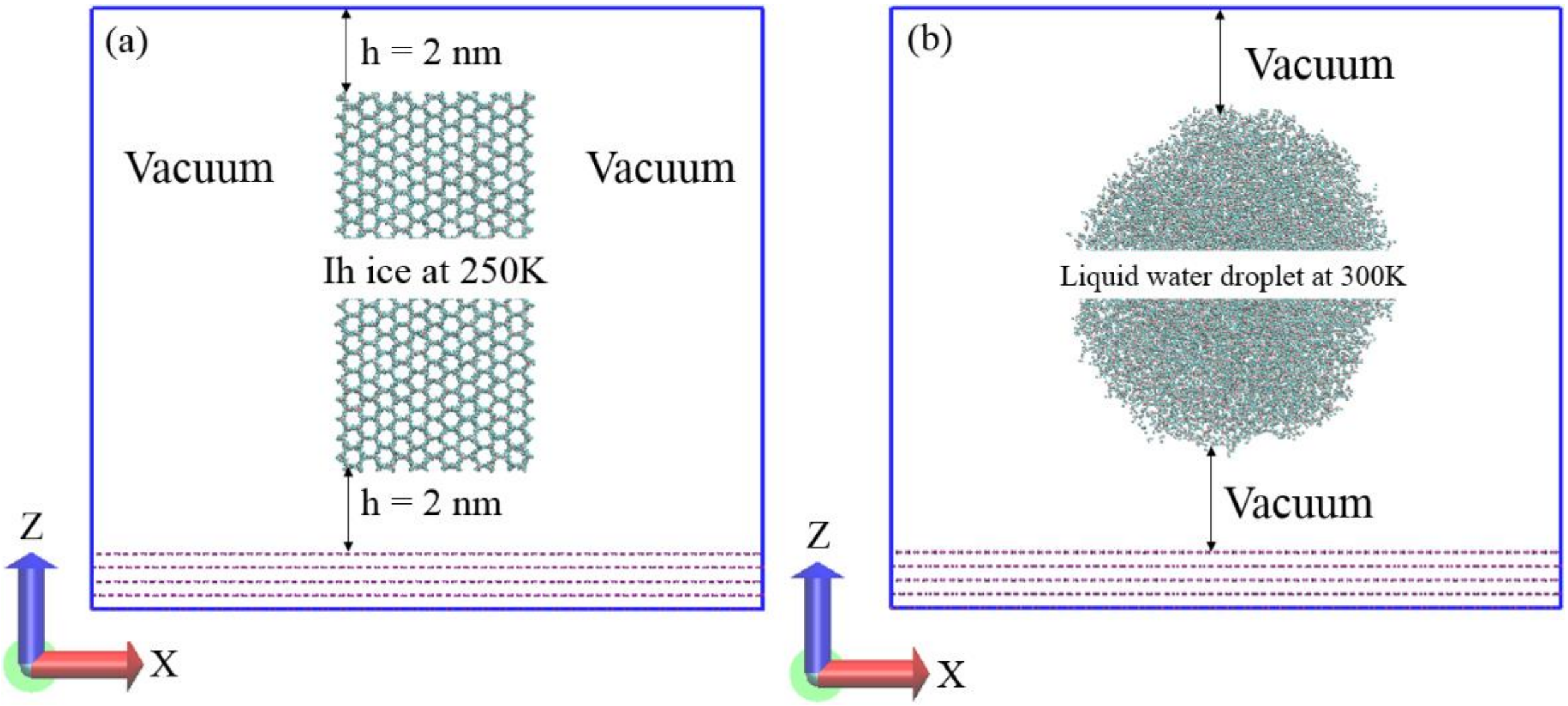

**Fig.1.** structure of water molecules on top of five graphene sheets; (a) Ih ice, (b) liquid water droplet at 300K.

In follows, to measure the equilibrium contact angle, the equilibrated liquid water droplet initially placed on the closest possible distance with graphene sheet where there is no atomistic overlap, through energy minimization process. Then the water droplet is equilibrated for 1.0ns under NVT simulation at 300K. In our study, interactions between water-substrate is calculated by setting force field parameters between oxygen (O) and carbon (C) atoms of graphene sheets. To model the van der Waals forces between water molecules and carbon atoms of graphene layers, we set the interaction parameters using the mixing rule we applied in our fracture study[22] and the values which are reported by Werder et. al.[6] Table 3 shows the force field interaction LJ parameters between ice and graphite substrate which governs the interaction potential of the system.

**Table 3.** Non-bonded LJ parameters between oxygen and carbon atoms

| | $\varepsilon_{O-C}$ **[kj mol-1]** | $\sigma_{O-C}$ [Å] |
|---|---|---|
| **Mixing rule** | 0.50840 | 3.3584 |
| **Werder et. al** | 0.39200 | 3.1900 |

Figure 2 illustrates the equilibrium states of liquid water droplet on five horizontal graphene sheets using different force field parameters, as reported in Table 3. Since it illustrates, the water equilibrium contact angle is between 50°-60° using the values from mixing rules. Experiments by Wang *et al*[28], and Shin *et al.*[29] indicate a water droplet equilibrium contact angle of 90-95° on graphite substrate, which is fairly well reproduced in our study by applying the force field parameters from the work of Werder et. al[6]. It should be noted that the equilibrium water droplet contact angle on graphene sheets strongly depends on the number of graphene sheets, size of the water droplet, atomistic water model and the system temperature[30]

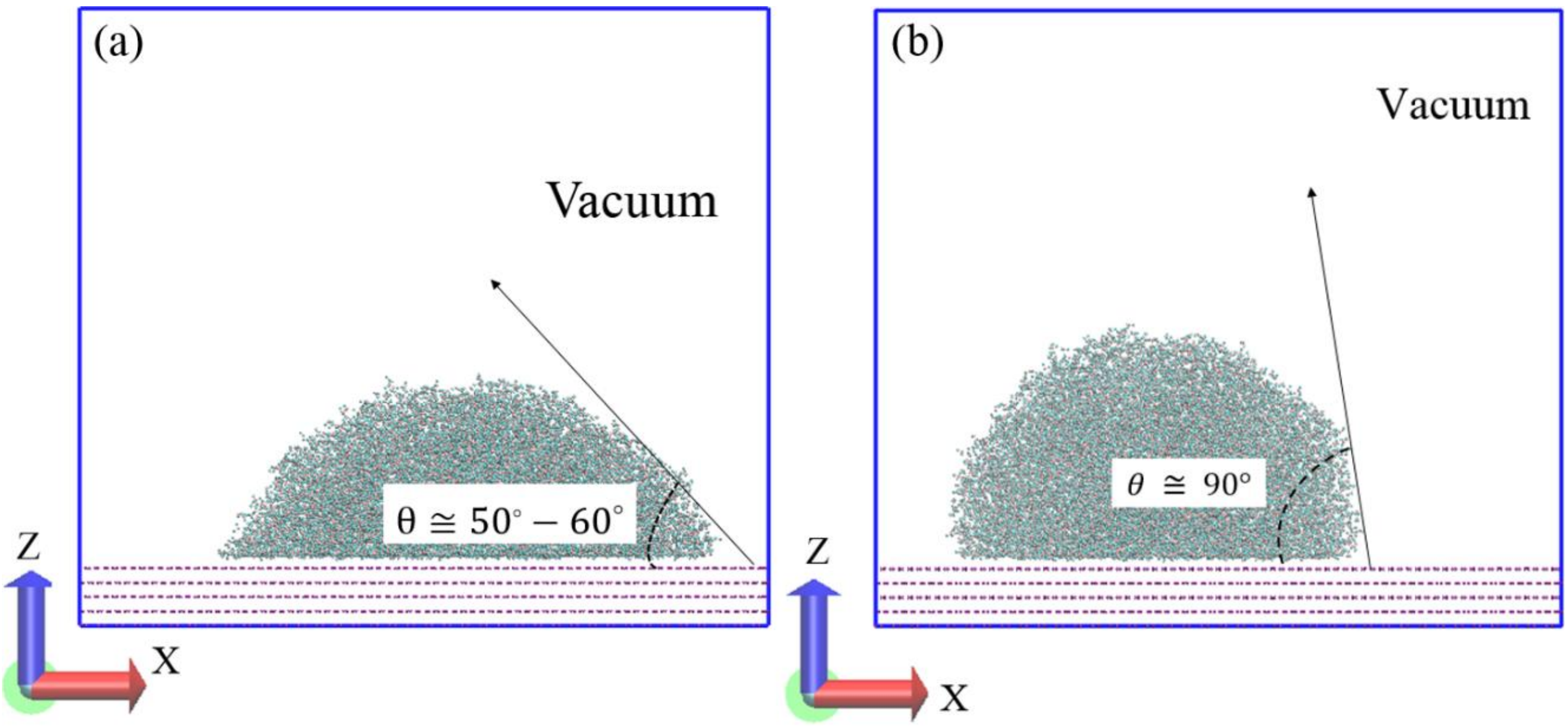

**Figure 2.** contact angle of liquid water droplet on graphene sheets using the force field parameters; (a) based on mixing rule, (b) work of Werder et al.[6]

Based on the obtained equilibrium contact angle of liquid water droplet on graphene sheets through using the developed force field parameters, we designed our simulation setup to study the wetting behavior of impinging supercooled liquid water droplet on nanostructure hydrophobic graphite substrate. To investigate the wetting profile of supercooled liquid water droplet, we initially placed prismatic facet of pre-equilibrated ice *I*h in a distance 2nm perpendicular to the graphene pillars. To ensure that the atoms do not interact with their periodic images at any time during the simulation, x-dimension of simulation box is increased. The ice cube was supercooled at two supercooling temperatures of 250K and 265K. In our study, graphite substrate built by several zig-zag edge type of graphene pillars each having height and width of 5.0nm and 6.6nm respectively along 'x' and 'y' directions which are extended along 'z' direction by 3.35Å[31].

We introduced groove structures to the graphite substrate by making some distances between graphite blocks called as "d" value. Based on the value of "d" parameter of 2nm and 4.7nm in this study, number of graphene pillars in each graphite block varied from 22, at d=2nm, to 14 sheets at d=4.7nm. Figure 3 illustrates a molecular schematic of the constructed system consists of pre-equilibrated ice and supercooled liquid water droplet at 250K on top of graphite substrate having 'd' equals to 2nm and 4.7nm

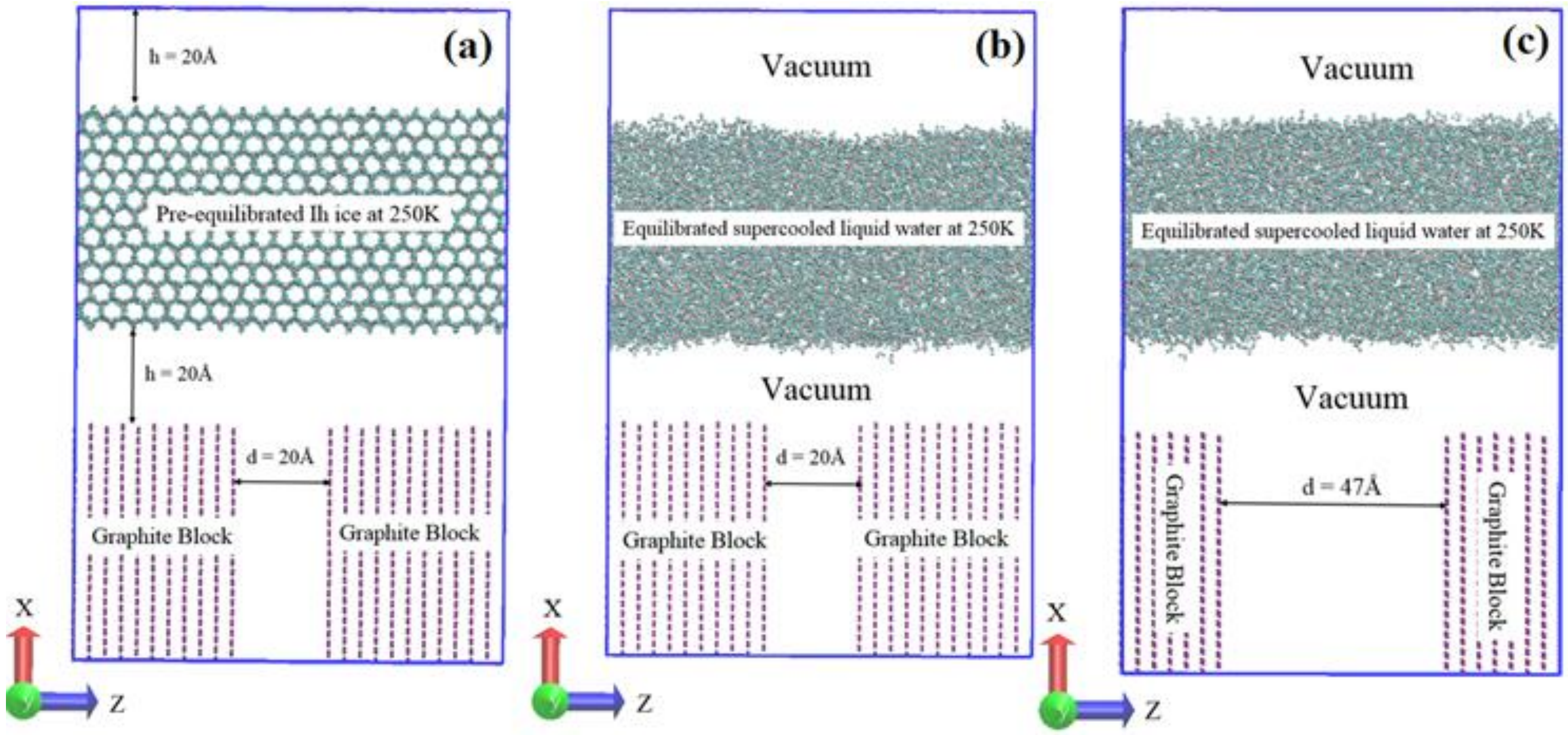

**Figure 3.** Molecular representation of the constructed model system for wetting simulation; (a) initial system with introducing Ih ice on top of graphite substrate, (b) supercooled liquid water at 250K on top of graphite blocks with 'd' at 2.0nm, (c) supercooled liquid water at 250K on top of graphite blocks with 'd' at 4.7nm

To investigate the effect of impinging velocity on wetting profile of supercooled liquid water droplet, we studied two different systems of impinging state and reference state, where there is no impact velocity. In reference state, equilibrated supercooled liquid water droplet brings into contact with the graphite substrate through energy minimization simulations. In impinging state, compared to our previous study[32], we introduced collective impinging velocity in terms of Much numbers (M) to the thermal velocities of equilibrated supercooled liquid water in a range of 0.5M, 1.0M, 1.5M, 2.0M, and 2.5M. Simulations are all conducted under NVT constraint with periodic boundary conditions in all directions of simulation box for 10ns with timestep of 0.001ps.

## 3. Results and Discussions

### 3.1. wetting profiles of water droplet

Figure 4 shows a molecular illustration of impinged supercooled liquid water with impact velocity of 0.5M and 2.5M at droplet temperature of 250K on graphite substrates having different surface roughness of 2.0nm and 4.7nm.

To calculate the depth of supercooled water penetration (H) into surface groove nanostructure, we used the developed formula as:

$$H \equiv \frac{2\int_{x=0}^{x=lg} x\rho(x)dx}{\int_{x=0}^{x=lg} \rho(x)\, dx} \qquad (4)$$

In this expression, $lg$ is the length of single graphene sheet, and $\rho(x)$ is the density of penetrated water molecules (specifically oxygen atoms) confined by graphene sheets which is varying during the simulation. This is calculated by the given equation 5, in which boundaries along 'z' direction is equal to the size of the surface groove structure in each substrate and $\rho_O$ stands for density of oxygen atoms of water molecules.

$$\rho(x) = \frac{1}{L_y L_z}\left[\iint_{ylo}^{yhi} \rho_O(x,y,z)\,dydz\right] \tag{5}$$

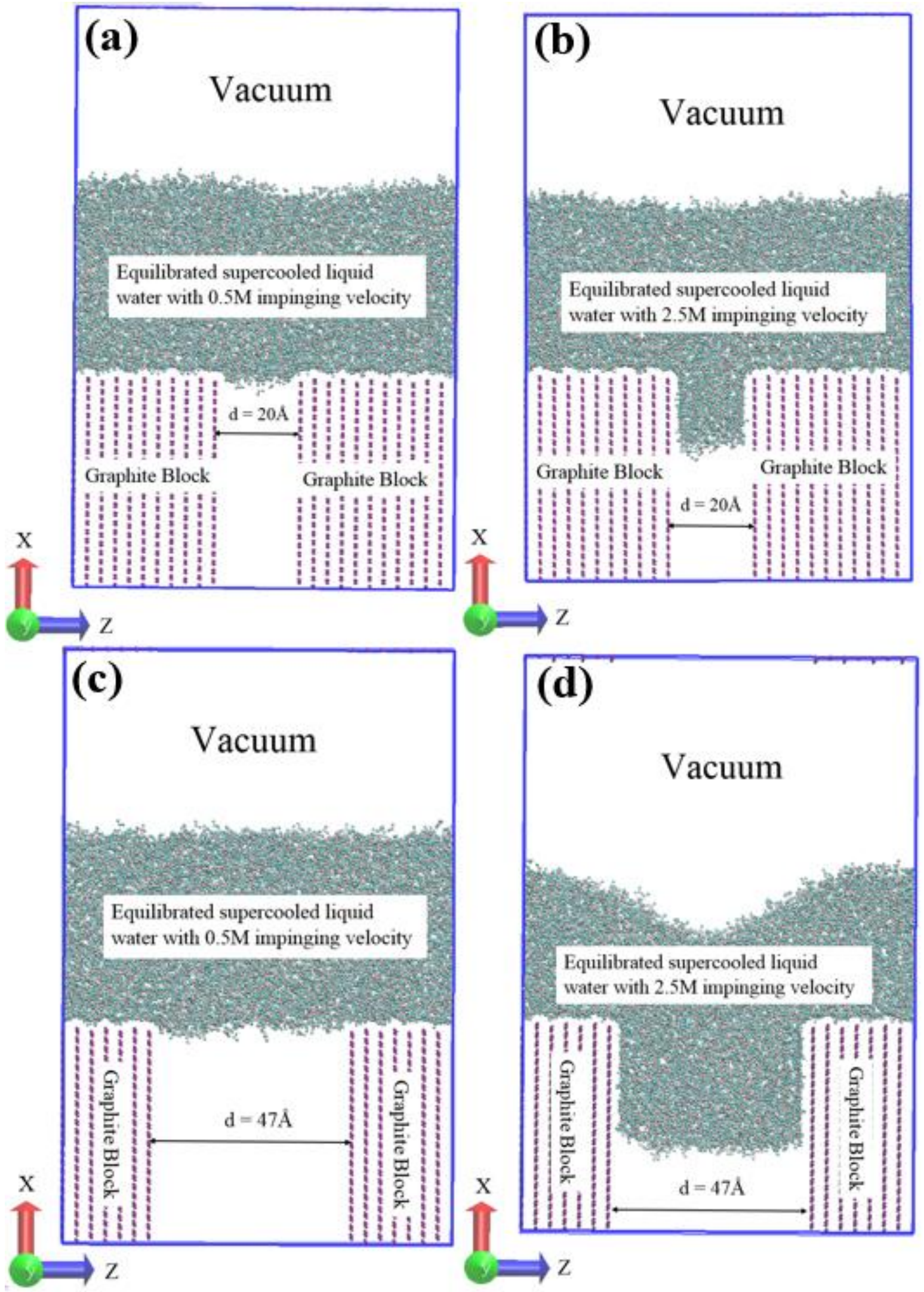

**Figure 4**. schematic of wetting behavior of supercooled liquid water on top of graphite substrate at different "groove" sizes and impinging velocities at 0.2ns; (a) impinging velocity of 0.5M, d= 2.0nm, (b) impinging velocity of 2.5M, d= 2.0nm, (c) impinging velocity of 0.5M, d = 4.7nm, (d) impinging velocity of 2.5M, d = 4.7nm.

Figure 5 indicates the evolution of depth of penetration for supercooled liquid water during wetting simulation at different droplet temperatures of 250K and 265K with d=2.0nm and 4.7nm. It shows supercooled water droplet demonstrates a gradual retreating behavior from surface texture structure once it penetrates through the substrate. This reflects the inherent hydrophobicity of graphite substrate as we modeled by calibrating the water-graphene interaction strength ($\varepsilon_{co}$). Since it demonstrates, retreating behavior is substantially depend on droplet impinging velocity and the size of surface roughness parameter. Also results shows no significant different in wetting profile of supercooled water droplet having impinging velocity equal and less than 1.0M. In this study, we define impinging velocity of 1.5M as the critical impinging velocity that results in fundamental change in wetting behavior of supercooled droplet on graphite substrate. Having this in mind, water droplet having higher impinging velocity shows higher depth penetration on surface having bigger texture size, which is true at different supercooling temperature. To compare wetting dynamics of impinged supercooled water with static wetting, where there is no impact droplet velocity, we conducted simulations for static supercooled liquid water droplets on graphite substrate at different temperatures of 250K and 265K and provided the wetting profile by the black ‘triangle’ data points in this figure. Comparing the dynamic with static wetting profiles, it indicates similar wetting profile for impinged water droplet after some simulation times called as ‘retracting time’ which is strongly depend on the droplet and surface physical properties.

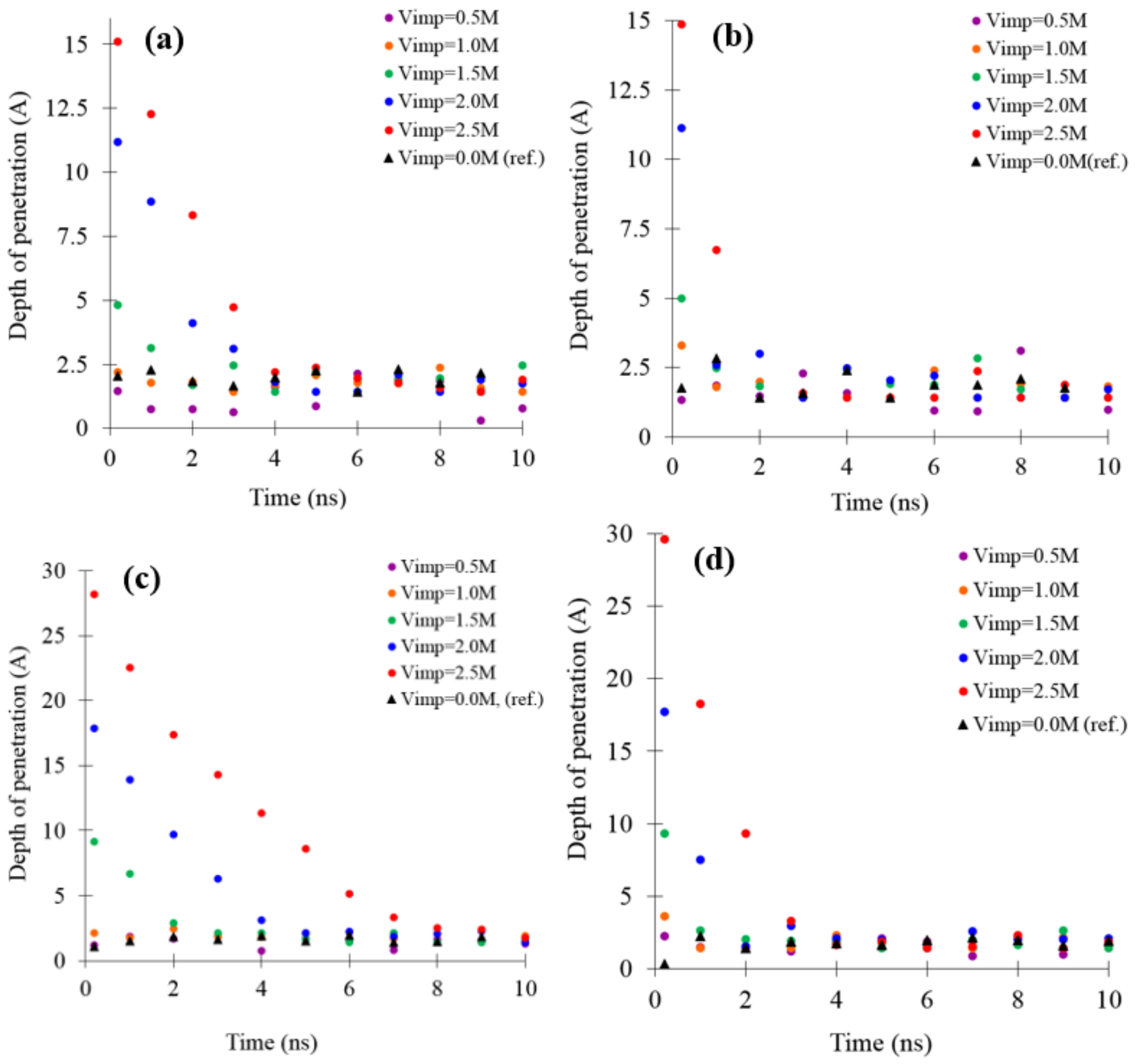

**Figure 5.** Wetting profiles of supercooled liquid water as a function of simulation time for (a) d=20Å, T = 250K, (b) d = 20Å, T=265K, (c) d=47Å, T = 250K, (d) d=47Å, T = 265K.

### 3.2. Initial penetration and retracting time profiles of water droplet

Figure 6 illustrates the initial penetration of supercooled liquid water as a function of impinging velocity and retracting time (ns) versus initial penetration. In these plots, we define the maximum H value represented in figure 5 as the initial penetration and retracting time as the time required dynamic wetting indicates similar static wetting.

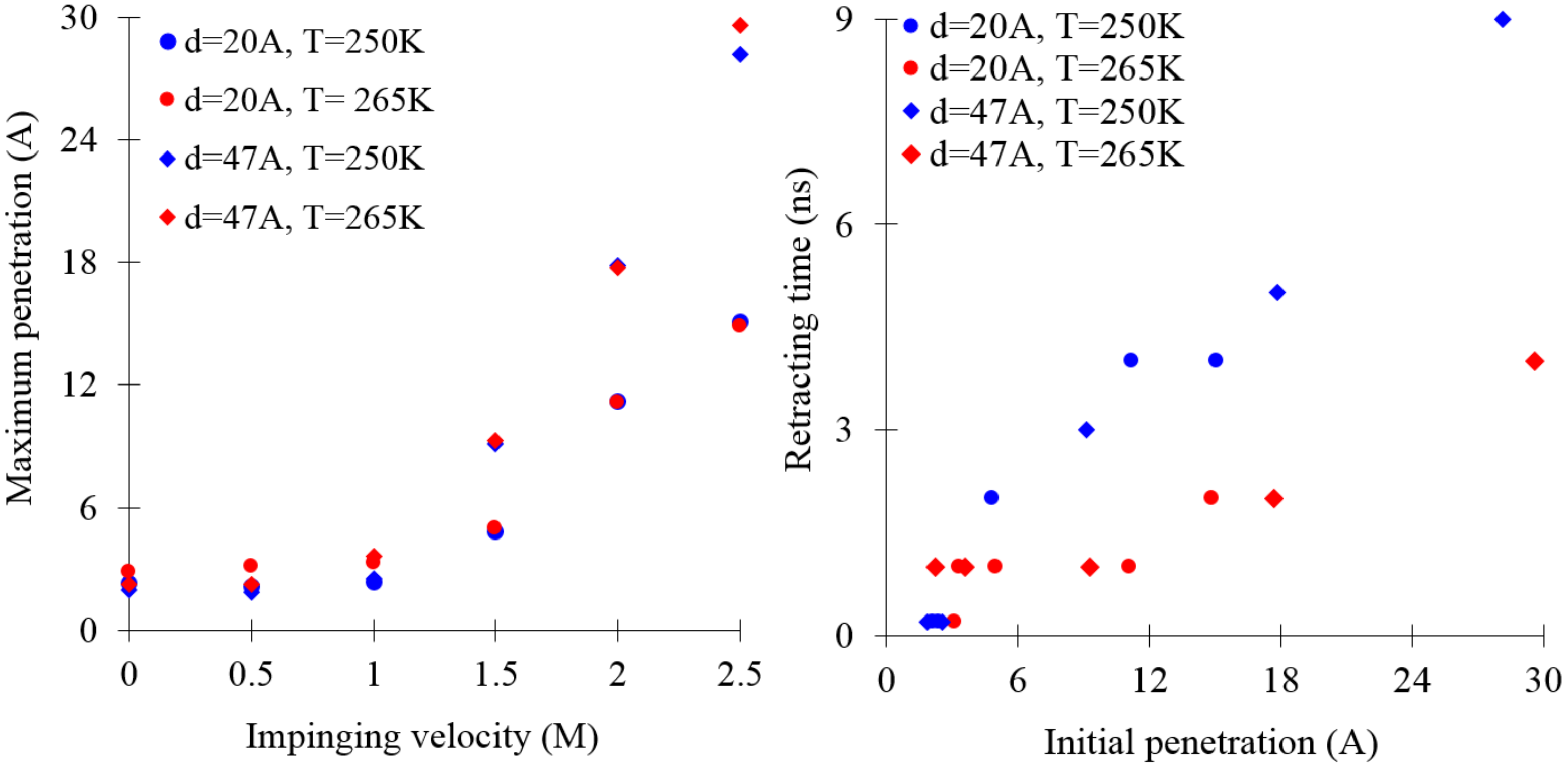

**Figure 6.** Initial penetration vs. imping velocity and retracting time vs. initial penetration.

These results demonstrate maximum penetration substantially depends on size of surface roughness and retracting time is significantly influenced by the droplet temperature. Results of maximum penetration vs. impinging velocity verifies the results shown in figure 5. It indicates droplet temperature and surface roughness initiates to indicate their influence on maximum depth of water penetration (MDP) starting at 1.5M in which MDP is noticeably increased by mounting impinging velocity and surface roughness. Result of retracting time represents supercooled water droplets requires higher time to be retracted from groove structure at lower temperature of 250K. We can address this to the potential of the TIP4P/ice atomistic water model we used to indicate the effect of temperature on viscosity of supercooled water droplet. Also, it illustrates retracting time is a strongly depend on droplet's temperature and the surface texture size. Figure 7 shows droplet retracting time as a function of impinging velocity. It demonstrates results shown in figure 7. Considering 1.5M as the critical impinging velocity, it verifies the dominant effect of droplet temperature on retracting profile of supercooled water from surface nanotexture structure. It indicates water impinging velocity has a considerable effect on retracting profile of water droplets having lower temperature. We can use all this useful information at atomistic scale in designing anti-icing surface at macroscale level.

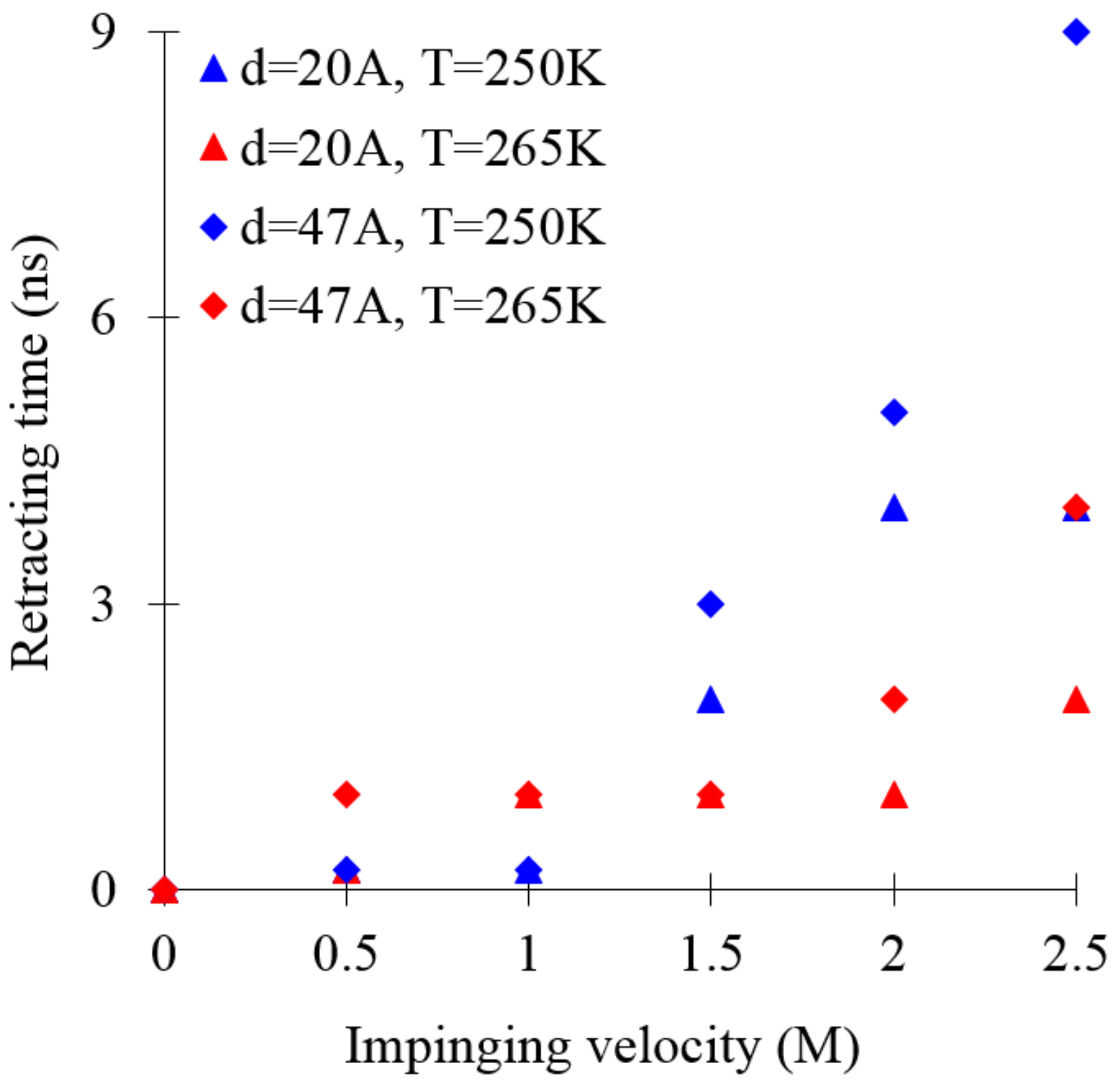

**Figure 7.** Retracting time vs. supercooled impact velocity for different surface "groove" sizes and droplet's temperature.

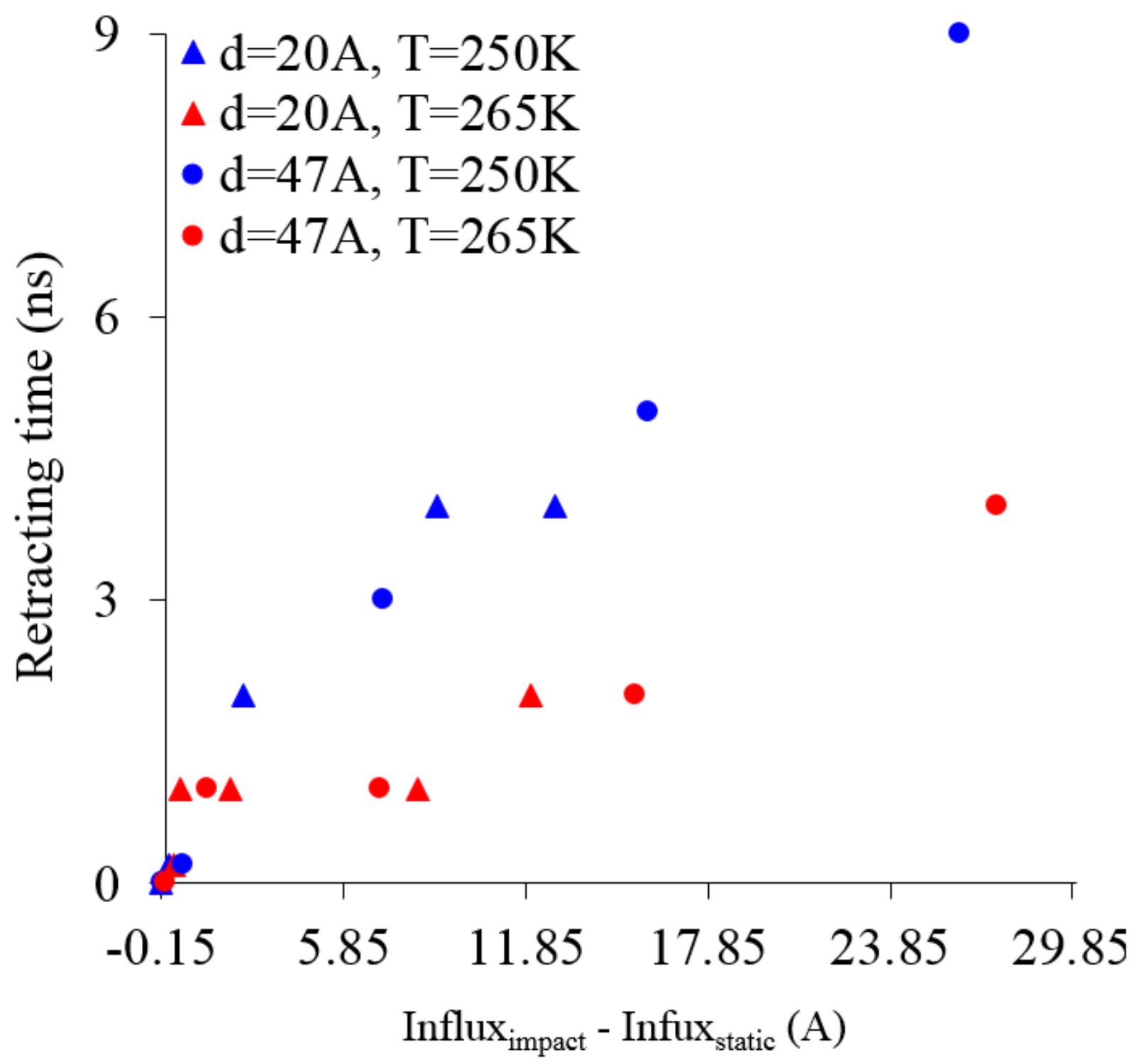

**Figure 8.** Retracting time as a function of initial penetration of impinged water droplet relative to the non-impinged droplet.

Figure 8 illustrates the retracting as a function of "pure" depth of penetration, where we subtracted the MDP values from the non-impacted water droplet. Result demonstrates the strong effect of supercooled water droplet's temperature on retracting time, which is remarkably higher at 250K compared to 265K at two 'd' values. This supports the results illustrated in figure 8. In detail, it demonstrates at a specific "d" value, supercooled water droplet with lower temperature requires more time to retract from surface groove structure.